# APPLICATION OF DATA MINING IN BIOINFORMATICS

KHALID RAZA

*Centre for Theoretical Physics,*
*Jamia Millia Islamia, New Delhi-110025, India*

**Abstract**

This article highlights some of the basic concepts of bioinformatics and data mining. The major research areas of bioinformatics are highlighted. The application of data mining in the domain of bioinformatics is explained. It also highlights some of the current challenges and opportunities of data mining in bioinformatics.

*Keywords*: Data Mining, Bioinformatics, Protein Sequences Analysis, Bioinformatics Tools.

## 1. Introduction

In recent years, rapid developments in genomics and proteomics have generated a large amount of biological data. Drawing conclusions from these data requires sophisticated computational analyses. Bioinformatics, or computational biology, is the interdisciplinary science of interpreting biological data using information technology and computer science. The importance of this new field of inquiry will grow as we continue to generate and integrate large quantities of genomic, proteomic, and other data.

A particular active area of research in bioinformatics is the application and development of data mining techniques to solve biological problems. Analyzing large biological data sets requires making sense of the data by inferring structure or generalizations from the data. Examples of this type of analysis include protein structure prediction, gene classification, cancer classification based on microarray data, clustering of gene expression data, statistical modeling of protein-protein interaction, etc. Therefore, we see a great potential to increase the interaction between data mining and bioinformatics.

## 2. Bioinformatics

The term bioinformatics was coined by Paulien Hogeweg in 1979 for the study of informatic processes in biotic systems. It was primary used since late 1980s has been in genomics and genetics, particularly in those areas of genomics involving large-scale DNA sequencing.

Bioinformatics can be defined as the application of computer technology to the management of biological information. Bioinformatics is the science of storing, extracting, organizing, analyzing, interpreting and utilizing information from biological sequences and molecules. It has been mainly fueled by advances in DNA sequencing and mapping techniques. Over the past few decades rapid developments in genomic and other molecular research technologies and developments in information technologies have combined to produce a tremendous amount of information related to molecular biology. The primary goal of bioinformatics is to increase the understanding of biological processes.

Some of the grand area of research in bioinformatics includes:

### 2.1. *Sequence analysis*

Sequence analysis is the most primitive operation in computational biology. This operation consists of finding which part of the biological sequences are alike and which part differs during medical analysis and genome mapping processes. The sequence analysis implies subjecting a DNA or peptide sequence to sequence alignment, sequence databases, repeated sequence searches, or other bioinformatics methods on a computer.

## 2.2. *Genome annotation*

In the context of genomics, annotation is the process of marking the genes and other biological features in a DNA sequence. The first genome annotation software system was designed in 1995 by Dr. Owen White.

## 2.3. *Analysis of gene expression*

The expression of many genes can be determined by measuring mRNA levels with various techniques such as microarrays, expressed cDNA sequence tag (EST) sequencing, serial analysis of gene expression (SAGE) tag sequencing, massively parallel signature sequencing (MPSS), or various applications of multiplexed in-situ hybridization etc. All of these techniques are extremely noise-prone and subject to bias in the biological measurement. Here the major research area involves developing statistical tools to separate signal from noise in high-throughput gene expression studies.

## 2.4. *Analysis of protein expression*

Gene expression is measured in many ways including mRNA and protein expression, however protein expression is one of the best clues of actual gene activity since proteins are usually final catalysts of cell activity. Protein microarrays and high throughput (HT) mass spectrometry (MS) can provide a snapshot of the proteins present in a biological sample. Bioinformatics is very much involved in making sense of protein microarray and HT MS data.

## 2.5. *Analysis of mutations in cancer*

In cancer, the genomes of affected cells are rearranged in complex or even unpredictable ways. Massive sequencing efforts are used to identify previously unknown point mutations in a variety of genes in cancer. Bioinformaticians continue to produce specialized automated systems to manage the sheer volume of sequence data produced, and they create new algorithms and software to compare the sequencing results to the growing collection of human genome sequences and germline polymorphisms. New physical detection technologies are employed, such as oligonucleotide microarrays to identify chromosomal gains and losses and single-nucleotide polymorphism arrays to detect known point mutations. Another type of data that requires novel informatics development is the analysis of lesions found to be recurrent among many tumors.

## 2.6. *Protein structure prediction*

The amino acid sequence of a protein (so-called, primary structure) can be easily determined from the sequence on the gene that codes for it. In most of the cases, this primary structure uniquely determines a structure in its native environment. Knowledge of this structure is vital in understanding the function of the protein. For lack of better terms, structural information is usually classified as secondary, tertiary and quaternary structure. Protein structure prediction is one of the most important for drug design and the design of novel enzymes. A general solution to such predictions remains an open problem for the researchers.

## 2.7. *Comparative genomics*

Comparative genomics is the study of the relationship of genome structure and function across different biological species. Gene finding is an important application of comparative genomics, as is discovery of new, non-coding functional elements of the genome. Comparative genomics exploits both similarities and differences in the proteins, RNA, and regulatory regions of different organisms. Computational approaches to genome comparison have recently become a common research topic in computer science.

## 2.8. *Modeling biological systems*

Modeling biological systems is a significant task of systems biology and mathematical biology. Computational systems biology aims to develop and use efficient algorithms, data structures, visualization and communication tools for the integration of large quantities of biological data with the goal of computer modeling. It involves the use of computer simulations of biological systems, like cellular subsystems such as the networks of metabolites and enzymes, signal transduction pathways and gene regulatory networks to both analyze and visualize the complex connections of these cellular processes. Artificial life is an attempt to understand evolutionary processes via the computer simulation of simple life forms

### 2.9. *High-throughput image analysis*

Computational technologies are used to accelerate or fully automate the processing, quantification and analysis of large amounts of high-information-content biomedical images. Modern image analysis systems augment an observer's ability to make measurements from a large or complex set of images. A fully developed analysis system may completely replace the observer. Biomedical imaging is becoming more important for both diagnostics and research. Some of the examples of research in this area are: clinical image analysis and visualization, inferring clone overlaps in DNA mapping, Bioimage informatics, etc.

### 2.10. *Protein-protein docking*

In the last two decades, tens of thousands of protein three-dimensional structures have been determined by X-ray crystallography and Protein nuclear magnetic resonance spectroscopy (protein NMR). One central question for the biological scientist is whether it is practical to predict possible protein-protein interactions only based on these 3D shapes, without doing protein-protein interaction experiments. A variety of methods have been developed to tackle the Protein-protein docking problem, though it seems that there is still much work to be done in this field.

## 3. Bioinformatics Tools

Following are the some of the important tools for bioinformatics (Table 1)

Table 1: Some of the tools for bioinformatics

| Bioinformatics Research Area | Tool (Application) | References |
|---|---|---|
| Sequence alignment | BLAST | http://blast.ncbi.nlm.nih.gov/Blast.cgi |
| | CS-BLAST | ftp://toolkit.lmb.uni-muenchen.de/csblast/ |
| | HMMER | http://hmmer.janelia.org/ |
| | FASTA | www.ebi.ac.uk/fasta33 |
| Multiple sequence alignment | MSAProbs | http://msaprobs.sourceforge.net/ |
| | DNA Alignment | http://www.fluxus-engineering.com/align.htm |
| | MultAlin | http://multalin.toulouse.inra.fr/multalin/multalin.html |
| | DiAlign | http://bibiserv.techfak.uni-bielefeld.de/dialign/ |
| Gene Finding | GenScan | genes.mit.edu/GENSCAN.html |
| | GenomeScan | http://genes.mit.edu/genomescan.html |
| | GeneMark | http://exon.biology.gatech.edu/ |
| Protein Domain Analysis | Pfam | http://pfam.sanger.ac.uk/ |
| | BLOCKS | http://blocks.fhcrc.org/ |
| | ProDom | http://prodom.prabi.fr/prodom/current/html/home.php |
| Pattern Identification | Gibbs Sampler | http://bayesweb.wadsworth.org/gibbs/gibbs.html |
| | AlignACE | http://atlas.med.harvard.edu/ |
| | MEME | http://meme.sdsc.edu/ |
| Genomic Analysis | SLAM | http://bio.math.berkeley.edu/slam/ |
| | Multiz | http://www.bx.psu.edu/miller_lab |
| Motif finding | MEME/MAST | http://meme.sdsc.edu |
| | eMOTIF | http://motif.stanford.edu |

## 4. Data Mining

Data mining refers to extracting or "mining" knowledge from large amounts of data. Data Mining (DM) is the science of finding new interesting patterns and relationship in huge amount of data. It is defined as "the process of discovering meaningful new correlations, patterns, and trends by digging into large amounts of data stored in warehouses". Data mining is also sometimes called Knowledge Discovery in Databases (KDD). Data mining is not specific to any industry. It requires intelligent technologies and the willingness to explore the possibility of hidden knowledge that resides in the data.

Data Mining approaches seem ideally suited for Bioinformatics, since it is data-rich, but lacks a comprehensive theory of life's organization at the molecular level. The extensive databases of biological information create both challenges and opportunities for development of novel KDD methods. Mining biological data helps to extract useful knowledge from massive datasets gathered in biology, and in other related life sciences areas such as medicine and neuroscience.

### 4.1. *Data mining tasks*

The two "high-level" primary goals of data mining, in practice, are prediction and description. The main tasks well-suited for data mining, all of which involves mining meaningful new patterns from the data, are:

*Classification:* Classification is learning a function that maps (classifies) a data item into one of several predefined classes.
*Estimation:* Given some input data, coming up with a value for some unknown continuous variable.
*Prediction:* Same as classification & estimation except that the records are classified according to some future behaviour or estimated future value).
*Association rules*: Determining which things go together, also called dependency modeling.
*Clustering:* Segmenting a population into a number of subgroups or clusters.
*Description & visualization:* Representing the data using visualization techniques.

Learning from data falls into two categories: directed ("supervised") and undirected ("unsupervised") learning. The first three tasks – classification, estimation and prediction – are examples of supervised learning. The next three tasks – association rules, clustering and description & visualization – are examples of unsupervised learning. In unsupervised learning, no variable is singled out as the target; the goal is to establish some relationship among all the variables. Unsupervised learning attempts to find patterns without the use of a particular target field.

The development of new data mining and knowledge discovery tools is a subject of active research. One motivation behind the development of these tools is their potential application in modern biology.

### 5. Application of Data Mining in Bioinformatics

Applications of data mining to bioinformatics include gene finding, protein function domain detection, function motif detection, protein function inference, disease diagnosis, disease prognosis, disease treatment optimization, protein and gene interaction network reconstruction, data cleansing, and protein sub-cellular location prediction.

For example, microarray technologies are used to predict a patient's outcome. On the basis of patients' genotypic microarray data, their survival time and risk of tumor metastasis or recurrence can be estimated. Machine learning can be used for peptide identification through mass spectroscopy. Correlation among fragment ions in a tandem mass spectrum is crucial in reducing stochastic mismatches for peptide identification by database searching. An efficient scoring algorithm that considers the correlative information in a tunable and comprehensive manner is highly desirable.

### 6. Conclusion and challenges

Bioinformatics and data mining are developing as interdisciplinary science. Data mining approaches seem ideally suited for bioinformatics, since bioinformatics is data-rich but lacks a comprehensive theory of life's organization at the molecular level.

However, data mining in bioinformatics is hampered by many facets of biological databases, including their size, number, diversity and the lack of a standard ontology to aid the querying of them as well as the heterogeneous data of the quality and provenance information they contain. Another problem is the range of levels the domains of expertise present amongst potential users, so it can be difficult for the database curators to provide access mechanism appropriate to all. The integration of biological databases is also a problem. Data mining and bioinformatics are fast growing research area today. It is important to examine what are the important research issues in bioinformatics and develop new data mining methods for scalable and effective analysis.